\documentclass[twocolumn]{el-author}

\usepackage{epstopdf}

\newcommand{\ua}{\uparrow}
\newcommand{\nc}{\newcommand}
\nc{\da}{\downarrow} \nc{\hc}{\hat{c}} \nc{\hS}{\hat{S}}
\nc{\bra}{\langle} \nc{\ket}{\rangle} \nc{\eq}{equation (\ref}
\nc{\h}{\hat} \nc{\hT}{\h{T}}\nc{\be}{\begin{eqnarray}}
\nc{\ee}{\end{eqnarray}}\nc{\rd}{\textrm{d}}\nc{\e}{eqnarray}\nc{\hR}{\hat{R}}\nc{\Tr}{\mathrm{Tr}}
\nc{\tS}{\tilde{S}}\nc{\tr}{\mathrm{tr}}\nc{\8}{\infty}\nc{\lgs}{\bra\ua,\phi|}\nc{\rgs}{|\ua,\phi\ket}
\nc{\hU}{\hat{U}}\nc{\lfs}{\bra\phi|}\nc{\rfs}{|\phi\ket}\nc{\hZ}{\hat{Z}}\nc{\hd}{\hat{d}}\nc{\mD}{\mathcal{D}}
\nc{\bd}{\bar{d}}\nc{\bc}{\bar{c}}\nc{\mc}{\mathcal}\nc{\ea}{eqnarray}\nc{\mG}{\mathcal{G}}\nc{\bce}{\begin{center}}
\nc{\ece}{\end{center}}
\date{9th March 2019}

\begin{document}

\title{Subframe resource optimization for massive machine device access in LTE networks}

\author{A. Ilori, A. Akindoyin, Z. Tang and J. He}

\abstract{\textbf{Abstract}: Synchronous massive machine device access can lead to severe congestion in the random access channel (RACH) of LTE networks. With scarce frequency resources, effective means must be developed to combat this key challenge. In this letter, the efficient allocation of frequency resources is considered as an optimization problem to be solved with a utility function. 
Based on this and a method of estimating the number of machine devices, an adaptive subframe allocation scheme is proposed. Numerical and simulation results verify the effectiveness of the proposed frame adaptation scheme in combating RACH congestion.}

\maketitle

\section{\textbf{Introduction}}
The ubiquity of cellular networks have made it an attractive option for emerging future communication paradigms. These advancements come with potential exponential increase in the number of devices needing network connectivity and would put a strain on the existing network, especially the RACH \cite{1}. Therefore, ways must be explored to cater for this projected increase especially in a way that guarantees satisfactory quality of service and throughput for both service providers and the end users amongst other metrics. This congestion in the RACH due to massive machine device access, is still an issue open to further research, with several solutions proposed such as \cite{2}.

In addition to the research efforts to improve the RACH access protocols, the RACH congestion problem can be alleviated by allocating more frequency resources (subframes) 
for machine devices communications, which is the subject of research in this letter.
According to the 3GPP specification there are 10 subframes in one frame. 
By default, 2 subframes are allocated for RACH 
while the remaining are used for data communications \cite{3}.
However, there is no specification from 3GPP or research reported on how to efficiently allocate the subframes for RACH 
to deal with the dynamic RACH loads while leaving sufficient resource for human devices. \\
In this paper, we propose an adaptive subframe allocation scheme to address the above problem.
The problem is formulated as an optimization problem with objective to maximize the system utility, 
which is defined as a function of RACH throughput and the subframe resource used by RACH.
We first solve the optimization problem to find the best configuration of the number of subframes for a known RACH load (i.e., number of machine devices).
Then we propose a method to estimate the RACH load according to the observations of historic RACH 
outcome at eNodeB to choose the number of subframes for the next round of random access.
This adaptive scheme produces an improvement in utility of about 75\% 
when compared to non-adaptive resource allocation schemes over the default RACH access protocol  \cite{4}.

\section{\textbf{Utility Function}}
The RACH is the first point of call for devices seeking to communicate in an LTE network. A number of steps, known as the RACH procedure, have to be successfully completed before devices are granted network access.  It involves a series of bidirectional information exchange between the eNodeB and the devices:
1) Preamble Selection: A unique selection of a preamble from a pool of 64 for RACH;
2) Network Response: appropriate messages are sent to the device notifying it of the success or otherwise of the previous step;
3) Contention Resolution: a back off procedure is implemented to resolve contention between devices.

Based on the RACH procedure, it is evident that a number of variables affect the RACH performance. Let $N_\text{d}$ denote the system load in terms of the number of devices seeking to access RACH. Also, let $N_\text{s}$ and $N_\text{p}$ denote the number of subframes and the number of preambles allocated for the RACH process respectively. 
We can set $N_\text{s}$ to a minimum value of 2 (the default) and a maximum of 8.  Considering a fixed system load, increasing $N_\text{s}$ will clearly have a positive effect on the number of devices that will be able to successfully complete the RACH process, even as devices will have more unique transmission opportunities, thus, having a positive effect on the system throughput, which is defined as the number of successful RA attempts measured in devices per frame. Clearly,  the performance of any system, in terms of throughput, is affected by two main factors, which are,  the number of devices seeking network access, $N_\text{d}$, and the number of subframes, $N_\text{s}$. 
Based on the RACH procedure, it is evident that a number of variables affect the RACH performance. Let $N_\text{d}$ denote the system load in terms of the number of devices seeking to access RACH. Also, let $N_\text{s}$ and $N_\text{p}$ denote the number of subframes and the number of preambles allocated for the RACH process respectively. 
We can set $N_\text{s}$ to a minimum value of 2 (the default) and a maximum of 8.  Considering a fixed system load, increasing $N_\text{s}$ will clearly have a positive effect on the number of devices that will be able to successfully complete the RACH process, even as devices will have more unique transmission opportunities, thus, having a positive effect on the system throughput, which is defined as the number of successful RA attempts measured in devices per frame. Clearly,  the performance of any system, in terms of throughput, is affected by two main factors, which are,  the number of devices seeking network access, $N_\text{d}$, and the number of subframes, $N_\text{s}$. 

Let $\eta$ denote the actual number of successful devices per frame, measured in devices per frame and let $\alpha$ be the benchmark number of devices per subframe. $\alpha$ is a variable under the control of the network operator , measured in device per subframe, that  regulates the RACH procedure within a subframe. The effects of $\alpha$ will be discussed further in the following section. Considering these parameters, $\eta$, $\alpha$ and $N_\text{s}$, we can derive a utility function with the objective of determining the ideal number of subframes to be used to achieve maximum utility.

Consequently, we define a simple utility function to assist in the subframe allocation, which is shown in Equation \ref{eqn5.1} below:
\begin{equation}
\label{eqn5.1}
U= \eta - \alpha N_\text{s},
\end{equation}
where $U$ denotes the system utility. 
It should be noted that the system utility has the same unit as throughput, however, more variables that affect the RACH procedure have been put into place and in order to improve on system performance, solving this utility function will be crucial.

\section{\textbf{Optimization Problem and Solution}}
Based on the utility function, a simple optimization problem can be formulated as follows: \\
Given a RACH load ($N_\text{d}$), find the best allocation of subframes ($N_\text{s}$) to maximize the system utility $U$.
An approach to solving this optimization problem is discussed next.

The throughput, $\eta$, can be expressed as a function of $N_\text{d}$, $N_\text{s}$ and $N_\text{p}$  as shown in Equation \ref{eqn5.2} \cite{5}:
\begin{equation}
\label{eqn5.2}
\eta = N_\text{d}  \text{exp} (- \frac {N_\text{d}}  {N_\text{s} N_\text{p}}) .
\end{equation}
\noindent
Substituting (\ref{eqn5.2}) into (\ref{eqn5.1}) becomes
\begin{equation}
\label{eqn5.3}
U =  N_\text{d}  \text{exp} (- \frac {N_\text{d}}  {N_\text{s} N_\text{p}})  - \alpha N_\text{s}.
\end{equation}
Differentiating (\ref{eqn5.3}) produces:
\begin{equation}
\label{eqn5.4}
\frac {dU} {d N_\text{s} } = - \frac {N_\text{d}^2}{ N_\text{p} N_\text{s}^2} \text{exp} (- \frac {N_\text{d}}  {N_\text{s} N_\text{p}})  + \alpha 
\end{equation}

At maximum value, $dU/d N_\text{s}= 0$. Equation (\ref{eqn5.4}) could be solved to yield and obtain the theoretical optimal value $N_\text{s}^{\text{opt}}$ for $N_\text{s}$,

\begin{equation}
\label{eqn5.5}
%\left\{
%\begin{array}{l@{\quad}l}
N_\text{s}^{\text{opt}}  \approx   -  { 2N_\text{p} \text{F}_\text{w}(-N_\text{d} \alpha N_\text{p} /N_\text{d}^{2})^{1/2}/2 N_\text{p}}\\\
% \,.
% \end{array} \right.
\end{equation}

\noindent
where $\text{F}_\text{w}$ is the Lambert W function. 

\begin{figure}[!h]
	\centering{\includegraphics[width=3.0in]{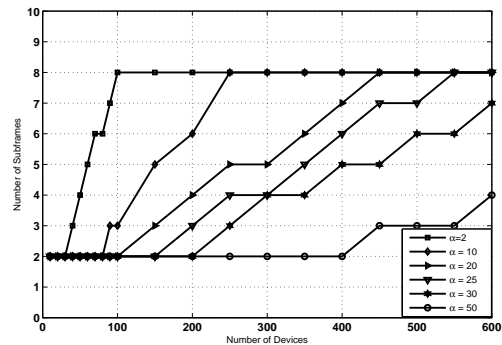}}
	\caption{Number of subframes under varying system load and  varying $\alpha$.}
	\label{SFvalues}
\end{figure}

According to the above formula, the number of subframes to be allocated, $N_\text{s}$, can easily be calculated for a given number of devices, $N_\text{d}$ and a fixed value of $\alpha$.
Figure \ref{SFvalues} shows the number of subframes against various loads with different values  $\alpha$.

The effect of $\alpha$ cannot be overemphasized, as it can be used by the network operator to regulate the allocation of subframes for random access. Note that a frame is divided into subframes, some of which are used to carry data whilst the remainder is used for the RACH procedure. Increasing or reducing one, has a direct effect on the other. 

From Figure \ref{SFvalues}, a low value of $\alpha$ quickly increases the number of subframes used for RACH procedure whilst a higher value gradually increases the number of subframes being allocated. For example, at $\alpha =2$, there is a sharp rise in number of subframes allocated, from $2$ subframes required for about 10 devices, to $6$ for $70$ devices, and then to the maximum of $8$ even at just $100$ devices. This is in contrast with 
an $\alpha$ value of $25$, which does not use the maximum number of subframes until about $550$ devices. It is also worthy of note that at $\alpha = 50$, we see the default number of subframes ($2$) being allocated until the system load reached $400$ devices, before a gradual increase afterwards. Consequently, at high values of $\alpha$, say $100$, the default number of subframes will be used to achieve maximum utility whilst at very low value or even zero value ($\alpha = 0$), the maximum number of subframes ($8$) will be required to achieve maximum utility irrespective of the number of devices. However, for the remainder of this work, we will use a value of $\alpha = 25$.\\
Consequently, as can be seen from Figure \ref{SFvalues}, we can easily determine the optimum number of subframes to be used by the network in providing effective service based on existing load. For values that are not within the range shown in the figure, we stick to the maximum number of subframes (8) in this case. In practise, based on the results obtained here, look up tables can be made offline and used to determine the optimum number of subframes for online operations.

Next, we consider the problem of estimating network loads by the eNodeBs, 
which is unknown but is required for the eNodeBs to determine the number of subframes$N_\text{s}$ for RACH as discussed before.
It is noted that the estimation can be performed on a very short time frame (e.g. for every frame period) to a long time frame (e.g. a few hours).
Due to the randomness of the network loads it is extremely difficult to accurately estimate the network loads.
In this work, we apply equation \ref{eqn5.2} to estimate the number of devices attempting to access the RACH for the last round of channel access.
As the eNodeB knows the throughput $\eta$ and the values of $N_\text{s}$ and $N_\text{p}$,
we can obtain an estimation of $N_\text{d}$ by solving equation \ref{eqn5.2} which has only one unknown variable.
It is noted that equation \ref{eqn5.2} can have two solutions for $N_\text{d}$ corresponding to the light and heavy network load conditions, respectively. 

\begin{equation}
\label{eqn5.6}
N_\text{d}  \approx \left\{
\begin{array}{l@{\quad}l}
-  {2 N_\text{p} N_\text{s} \text{F}_\text{w}(-(\alpha N_\text{p} N_\text{s}^2)^{1/2})/2 N_\text{p} N_\text{s}}\,;\\
-  {2 N_\text{p} N_\text{s} \text{F}_\text{w}((\alpha N_\text{p} N_\text{s}^2)^{1/2})/2 N_\text{p} N_\text{s}}\,.
\end{array} \right.
\end{equation}
where $\text{F}_\text{w}$ is the Lambert W function and $N_\text{p}$ is the number of preambles in an eNodeB.

In order to predict which of these is to be used, we use the preamble detection outcomes at the eNodeBs. 
The outcome of a preamble selection in the RACH process could be considered as one and only one of these: \\
1) a preamble is selected by a device and successfully decoded by the base station; \\
2) a preamble is picked by more than one device, in which case we assume there is a collision; \\
3) a preamble is not chosen by any device. \\
Unselected preambles could help distinguish when the system is under heavy load or not. 
Whilst the choice of preamble selection by devices is entirely random, we infer that if there is heavy load on the system, 
there would be fewer numbers of unselected preambles and vice versa. This helps estimate the numerical performance of the proposed scheme.
Then the eNodeB can use an average of the estimated number of devices as the estimated network load to be used for subframe allocation.

\section{\textbf{Performance Evaluation of the Subframe Resource Optimization Scheme}}
\begin{figure}[!tbh]
	\centering{\includegraphics[width=3.0in]{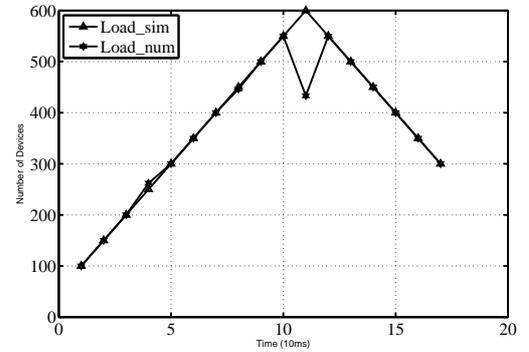}}
	\caption{Network load estimation over time.}
	\label{load}
\end{figure}

In this section, we evaluate the performance of the proposed subframe resource optimization scheme in terms of utility and system throughput. We also verify the reliability of the system load prediction equation proposed in \ref{eqn5.6} both by using simulations and numerically.
An homogeneous network with changing system load over time, served by a single eNodeB is considered. 
The arrival of  devices following a poisson distribution with the mean arrival rate increasing linearly first and then decreases linearly, as shown in Figure \ref{load}.
We assume that all $64$ preambles are available for use by the devices and simulate the RACH procedure with the assumption that the eNodeB has enough resources to accommodate devices that successfully complete the RACH procedure.
Performance metrics used in this study are the system throughput and the utility as obtained from (\ref{eqn5.3}). 

It should be recalled that the default RACH process only uses a fixed number of subframes (2), and is denoted by RACH in our figures. 

Next, we test the effectiveness of the method to estimate network loads by solving Equation \ref{eqn5.2}. Known values of network load, as displayed using Load\_sim in Figure \ref{load}, is used to predict the potential network load for the next round of access. For instance, under time $5$ms, the network load assumed under simulations is 300 devices, however, for numerical estimations, the last estimated value (number of devices estimated at $4$ms) is used to predict (using equation \ref{eqn5.2}) the potential number of devices for $5$ms and in this case, there is an exact match between the simulated and predicted values. However, cases exists where there is a mismatch between the simulated and numerical load values, (such as, at $11$ms) and this could be attributed to the way in which the system predicts whether the network is under heavy or light loads. As stated earlier, in our calculations, unselected preambles are used to distinguish the load condition of the network and this have proven to be accurate to a large extent, with more than $95\%$ of simulated values matching numerical calculations. Despite this rare mismatch occurrence, the proposal was able to adjust and correctly predict loads for the next round of random access and beyond, supporting the belief that this blip does not invalidate our proposed methodology. Conclusively, Figure \ref{load} shows a close match between the simulated and expected network load. Load\_sim and Load\_num are the values of load in simulation and numerical analysis respectively. This further lends credence to the derivation in Equation \ref{eqn5.6}.

\begin{figure}[!tbh]
	\centering{\includegraphics[width=3.0in]{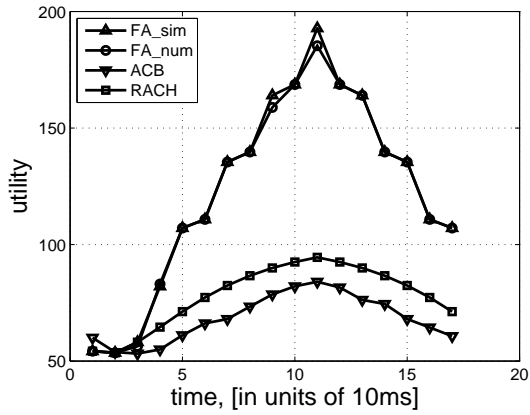}}
	\caption{Utility comparison of the fixed and adaptive subframe allocation scheme with time.}
	\label{utility}
\end{figure}

Fig.~\ref{utility} presents the utility obtained by the adaptive and fixed allocation schemes with the varying network loads.
Four performances are represented in the figure including, the utility performance of our proposed subframe adaptation scheme, both in simulation (FA\_sim) and by numerical analysis (FA\_num), the default RACH procedure and the ACB scheme.
Using network loads shown in Figure \ref{load}, we evaluate the performance of these schemes in terms of utility. Again, very close matches between utility values obtained via simulations and numerical calculations are observed, further supporting the validity of our derivations. In terms of performance, the benefit of the adaptive scheme is clearly evident as it outperforms both ACB and the default RACH process, with more than 100\% increase in utility in some instances. 
Both RACH and ACB are not able to obtain high values of utility and this will in effect translate to them not being able to cope with heavy network loads.\\

Further analysis of these schemes in terms of access delay, collision probability and throughput is carried out in the following sections. These metrics would be able to provide an interesting insight into the performances of these schemes and their impact on the QoS delivered to the network. 

However, as we have proven from Figure \ref{utility} and Equation~\ref{eqn5.1}, the cost at which an increase in throughput is provided might not always be optimal. 
In cases where all subframes are used to provide RA opportunities, the subframes for data are sacrificed. 
Thus, fixing the subframes to a maximum would be  detrimental to data services, 
whilst using the minimum number of subframes (RACH) would lead to degradable QoS under very heavy loads.

\section{\textbf{Conclusion}}
In conclusion, we proposed a frame adaptation scheme that is capable of handling massive machine device access in LTE networks. We define a utility function upon which an optimization problem is formulated and a solution is found to determine the optimal number of subframes to be allocated to the machine devices over RACH.
In addition, we proposed a method to estimate the number of devices attempting to access the RACH.
Simulation results have verified the effectiveness of our proposal over a fixed subframe allocation schemes such as the default RACH. 
Over 100\% increase in utility has been observed in some cases.

\vskip5pt

\noindent 

%\vskip3pt
\noindent Email:  ayoade.ilori@yahoo.com

\end{document}